\documentclass[prb,notitlepage,twocolumn]{revtex4}
\usepackage{amsmath}
\usepackage{amssymb}
\usepackage{bm}
\usepackage{graphicx}

\begin{document}

\title{Fano resonances in antennas: General control over radiation patterns}

\author{Mikhail V. Rybin${}^{1,2}$}
\email{m.rybin@mail.ioffe.ru}
\author{Polina V. Kapitanova${}^{2}$}
\author{Dmitry S. Filonov${}^2$}
\author{Alexey~P.~Slobozhanyuk${}^2$}
\author{Pavel A. Belov${}^2$}
\author{Yuri S. Kivshar${}^{2,3}$}
\author{Mikhail F. Limonov${}^{1,2}$}

\affiliation{$^1$ Ioffe Physical-Technical Institute, St.~Petersburg 194021, Russia\\
$^2$ National Research University for Information Technology, Mechanics and Optics (ITMO),
St.~Petersburg 197101, Russia,\\
$^3$Nonlinear Physics Center, Research School of Physics and Engineering,
Australian National University, Canberra ACT 0200, Australia
}

\begin{abstract}
The concepts of many optical devices are based on the fundamental physical phenomena such as resonances.
One of the commonly used devices is an electromagnetic antenna that converts localized energy
into freely propagating radiation and vise versa, offering unique capabilities for controlling electromagnetic
radiation. Here we propose {\em a concept} for controlling the intensity and directionality of
electromagnetic wave scattering in radio-frequency and optical antennas based on the physics of
Fano resonances. We develop an analytical theory of {\em spatial Fano resonances} in antennas that describes
switching of the radiation pattern between the forward and backward directions, and confirm our theory with
both numerical calculations and microwave experiments. Our approach bridges the concepts of conventional
radio antennas and photonic nanoantennas, and it provides a paradigm for the design of wireless
optical devices with various functionalities and architectures.
\end{abstract}

\date{\today}


\maketitle

\section{Introduction}

The communication flow transfused the modern world, and one of the key
elements of current communication networks is an antenna that
converts effectively the electromagnetic energy between localized and propagating waves.
Since the first design of antenna in 1886 by Heinrich Hertz,
 sizes, shapes, and materials of antennas were evolved dramatically
just reflecting both human needs and technology progress. Along with the technological
progress, a theory of antennas has been developed starting from the theory of
radio-frequency antennas formulated around the middle of XX century.

The first decade of the XXI century demonstrates the needs for
global miniaturization, and nanotechnology makes it possible to scale down many
electronic and photonic devices, thereby opening up new horizons for their applications.
One of such devices is an optical nanoantenna that converts localized energy
into freely propagating optical radiation and vise versa, offering unique capabilities
for controlling electromagnetic radiation at the nanoscale~\cite{biagioni2012nanoantennas}.
Nanoantennas are considered as key elements in nanodevices being designed for a wide
range of applications~\cite{novotny2011antennas,alu2008tuning,sivis2013extreme,barnard2011photocurrent,
galliker2012direct,knight2011photodetection,large2010photoconductively,
moerland2008reversible,yanik2011seeing}. Besides, recent evidence suggests that a variety
of organisms may harness an antenna for capturing sunlight and storing
its energy transiently\cite{lambert2012quantum,scholes2010biophysics}.

Recent research activities in nanophotonics are mainly focused on
the studies of plasmonic nanoantennas created by metal nanoparticles
that increase the effectiveness of radiation or reception of signals
of individual molecules or single quantum dots by several orders of
magnitude, due to the Purcell effect~\cite{Tame2013,bakker2008enhanced,kinkhabwala2009large,
kuhn2006enhancement, liu2011nanoantenna,taminiau2007lambda}. On the other hand,
an optical nanoantenna should provide a low level of dissipation losses that
is a serious weakness of all plasmonic devices. For this reason, several research
teams turn their attention to the study of all-dielectric antennas
with considerably low losses in the visible~\cite{filonov:201113,rolly2012boosting,fu2013directional}.

Efficiency any antenna can be enhanced substantially by resonance phenomena,
which can shape and govern the structure of far-field radiation patterns.
The resonances of different nature can be considered including the Bragg
resonance associated with periodically arranged elements, i.e. based
on a photonic crystal structure~\cite{g142}. However for problems addressed
by single quantum objects, the most fruitful approach is to lower
the antenna symmetry by adding some elements with resonance scattering.
In the radio-frequency range, such antennas corresponds to the concept proposed
by Yagi being based on a resonant element called {\em a reflector} that forms
a radiation pattern (the so-called Yagi-Uda antennas)~\cite{yagi1928beam}.
To obtain a narrow radiation pattern, several elements referred to as {\em directors}
are used additionally~\cite{yagi1928beam,g147}.

\begin{figure*}[!t]
\centerline{\mbox{\resizebox{14cm}{!}{\includegraphics{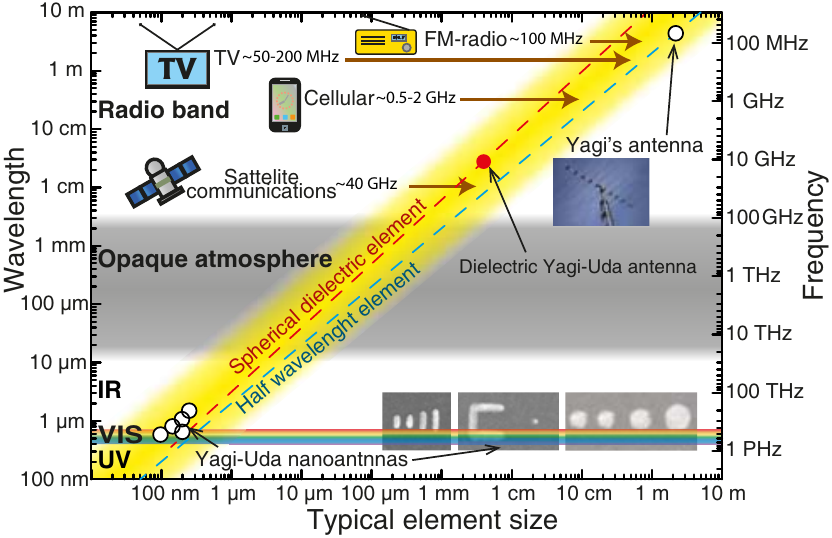}}}}
 \caption{ (Color online) World of antennas. Examples of different devices with
antennas and single antennas in the parameter plane antenna size vs. operating
frequency. Region of the typical size is marked by yellow. Circles show
the results of different experiments with Yagi-Uda radio-frequency and nanoantennas
from 
\cite{yagi1928beam,coenen2011directional,curto2010unidirectional,dorfmuller2011near,
dregely20113d,filonov:201113,galliker2012direct}. Red dot marks our experimental data.
}
\label{fig:Table}
\end{figure*}

Very different engineering solutions that define the antenna architecture can be reflected
in the electromagnetic spectrum shown in Fig.~\ref{fig:Table}. Here we refer to Yagi-Uda antennas
that have been studied for the optical frequency range with metallic nanorods
\cite{curto2010unidirectional,kosako2010directional,dorfmuller2011near,dregely20113d,
maksymov2011enhanced,novotny2011antennas,galliker2012direct,maksymov2012actively}.
Besides, the spherical metal particles and split-ring
resonators have been suggested as elements of Yagi-Uda antennas
\cite{li2007shaping,coenen2011directional,jie2013antenna,munarriz2013optical}.
We emphasize that all structure elements that used so far in the antenna design
are characterized by a resonant response to an external electromagnetic field
which, as we show below, can be properly engineered.


Here we propose a concept for controlling the intensity and directionality of electromagnetic
wave scattering for any antenna, including radio-frequency antennas and optical nanoantennas.
Our concept is based on the physics of Fano resonance~\cite{g704} that involves  both constructive
and destructive interference of a narrow frequency spectrum line with a broad spectral band.
If two scattering channels have different locations in space, the relative phase shift depends on
radiation directions that induce an extraordinary Fano resonance with the possibility of switching
between the forward and backward radiation directions. The Fano resonance is observed when the wave
scattering occurs through multiple channels, one of these channels possess a narrow resonance in which
the spectral dependence of phase undergoes a change by $\pi$, while other channels do not show a sharp
change in the phase, forming the so-called continuum. Fano resonance has been observed across many different
branches of physics, such as the studies of magnetization~\cite{g709} and electronic polarization
phenomena, semiconductor and superconductor~\cite{g706,limonov2002superconductivity}
optics, as well as in various nanoscale objects including photonic and plasmonic
structures~\cite{g901,g908,g020,g026,g030,francescato2012plasmonic,rahmani2012fano}.

We consider Fano resonance as an interference of two waves. The
first one is proportional to $Ae^{i\delta _{A}} (\varepsilon +i)^{-1}
$ (the Lorenz function) and the second one is defined as $Be^{i\delta
_{B} } $. Here $A$, $B$, and $\delta _{A,B} $ are real functions with
relative changes in the frequency range of interest being negligible
in comparison with the Lorenz function. Due to the interference, the
resulting wave intensity can be written in the form,
\begin{equation}\label{eq:FanoFormula}
 I(\omega )=\left(\frac{(q+\Omega )^{2} }{1+\Omega ^{2} } \eta
+(1-\eta )\right)B^{2},
\end{equation}
where $q$ is the Fano asymmetry parameter, $\Omega =(\omega -\omega
_{0} )/(\Gamma /2)$ is the dimensionless frequency, $\omega _{0} $ is
the central frequency, and $\Gamma $ is the width of the narrow band.
Experimentally observed background component that does not interfere
with the narrow band~\cite{g706,g020,g026} is
accounted by the introduction of an interaction coefficient $\eta \in
[0..1]$
\begin{equation}
\label{eq:eta}
\eta =\frac{2F\cos ^{2} \delta }{F+2\sin \delta +\sqrt{F^{2} +4F\sin
\delta +4} },
\end{equation}
where $F=A/B$ and $\delta =\delta _{A} -\delta _{B} $ are the relative
intensity and phase difference, correspondingly. The Fano parameter
of our special interest is
\begin{equation}\label{eq:FanoParameter}
q=\cos \delta \, F / \eta .
\end{equation}
Equation~(\ref{eq:FanoFormula}) shows that depending on the sign and value of $q$, the
Fano-type spectra have {\em four characteristic shapes}. For special values
of $q$ the narrow band is symmetrical ($q=0$ and $q>\pm \infty$)
while in the general case where $-\infty <q<0$ or $0<q<+\infty$, a
narrow band exhibits an asymmetric profile, that switching its shape
into mirror image with changing of the sign of $q$.

Here we demonstrate that the unique flexibility of the narrow band
leads naturally to a possibility for manipulating \emph{the spatial
dispersion of the antennas' radiation}. If resonant and nonresonant sources
are placed asymmetrically relative to the antenna center, the Fano resonance
will lead to the suppression of the radiation in some directions with
the radiation enhancement in other directions. As a result, an asymmetric antenna
with a resonance element will exhibit a dramatic modification of the radiation
patterns and switching of the main lobe between the forward and backward directions.
Here we consider a challenging case when the underlying mechanism of
the Fano resonance leads to the transformation of the radiation
pattern in the three-dimensional space while generally discussed Fano
analysis is carried out for the spectral scale
\cite{g901,francescato2012plasmonic,g908,rahmani2012fano}.
\emph{The Fano parameter $q$, which determines the shape
of the narrow line, in the case of antennas becomes spatially
dependent}. We call the antenna those operation is based on
the physics of Fano resonances as \emph{Fano-antenna}.

\section{Results}
\subsection{General approach based on the concept of Fano resonances}

First, we notice that for most types of antennas we can introduce
a source of the radiation characterized by a dipole moment. Because
of reciprocity, the source is usable for both emitting and absorbing
radiation, and hereafter it is referred to as E/A-feed. In
an antenna, the electromagnetic wave undergoes a series of scattering
events on the additional elements as well as on the E/A-feed that
leads to a change in the dipole moment of the E/A-feed and in the
response of the scatterers. A change of the dipole moment in the
presence of a resonator is known as the Purcell effect, which depends
on the geometry of the E/A-feed. We do not consider a specific type of
the E/A-feed, and normalize the dipole moment to unity. This
normalization has no effect on the relative amplitude of the radiation
propagating in different directions, i.e. on the far-field pattern.


Next, we introduce the second antenna element (director or reflector)
having a resonant response (hereafter referred to as D/R-element). As
a result, the electromagnetic wave may reach the same far-field state
via two different channels. The first traveling path from the E/A-feed
directly to the far-field zone corresponds to the formation of a broad
background, where the wave phase and amplitude are nearly constant in
the spectrum range of interest. The second traveling path includes the
scattering event on a D/R-element having a resonant response and leads
the formation of a narrow band, where the wave phase changes by $\pi$.
If the amplitude of the narrow band can be approximated by Lorenz
function, the result of the interference of two waves in the far-field
zone can be described by the Fano formula~(\ref{eq:FanoFormula}).

Departing from this general view and with the intention to further
deepen our understanding of emitting and absorbing of electromagnetic
waves in Fano-antennas, a number of challenging problems can be
formulated: How to separate a Lorenz function for the amplitude of the
narrow band for different real types of antennas? What determines the
Fano asymmetry parameter $q$? And probably the most important
question - is it possible to switch the direction of the main lobe by
changing the sign of the Fano asymmetry parameter $q$ at
realistic geometric parameters of antennas?

In general, a real D/R-element can not be described by only one
resonance and there is a superposition of multipole moments. We will
demonstrate that in most cases it is possible to separate an isolated
dipole moment, which in the resonance region can be described by the
Lorenz function multiplied by a slowly varying background originating
from other multipole moments. Using an analytical model presented
here, an optimal configuration of the Fano-antennas elements can be
found in a first approximation and parameters of the model can be
fitted numerically using a non-linear optimization method (e.g.
Levenberg-Marquardt method). Note that the interaction between the
E/A-feed and the D/R-element is taken into account by normalizing the
E/A-feed dipole moment.

The most detailed theory has been developed for radio antennas.
Traditional assumption about the description of radio antennas is
based on the framework of effective electrical parameters, which
depend on geometrical configuration of the antenna \cite{g147}. The
resonant frequency of the thin metal wire antenna corresponds to a
first approximation to the half-wave antenna condition $L\approx
\lambda /2$, where $L$ is a length of the wire. For the long
wire $L>\lambda /2$, the impedance has mainly inductive character, and
for the short wire $L<\lambda /2$ the impedance has capacitive
character. By contrast, here we consider the response of the antenna
to the electric field, which in the case $L<\lambda $ can be described
in terms of electric dipole momentum \cite{g143}.

\begin{figure}[t]
\includegraphics{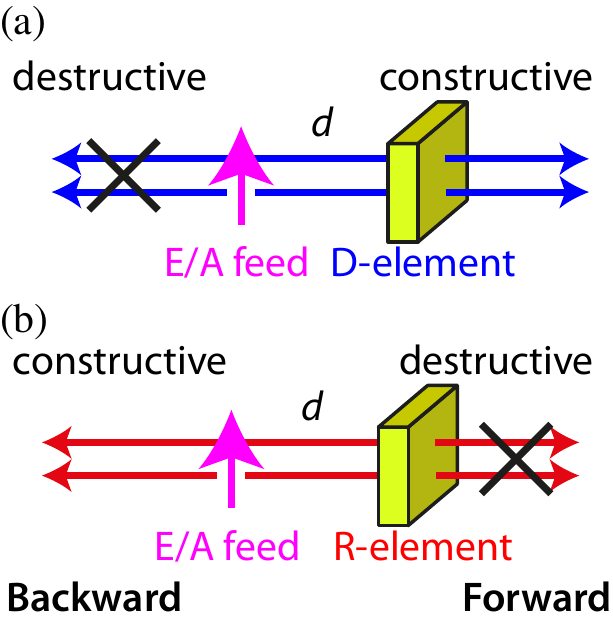}
 \caption{ (Color online) 
Schematic of the two regimes of Fano-antennas. The antenna is
composed of two elements: the emitting/absorbing element (E/A-feed)
and directing/reflecting element (D/R -- element). The upper diagram shows
the directing interference regime and the lower diagram shows the reflecting
interference regime. The distance between the E/A-feed and D/R -- element denoted by $d$.}
\label{fig:Scheme}
\end{figure}

To make the discussion more clear, we consider two regimes of the Fano-antenna
operation depicted schematically in Fig.~\ref{fig:Scheme}. A resonant D/R -- element
can have reflecting or directing properties depending on the phase of
the response. Note that the condition for a wave propagation to the
forward direction depends only on a phase shift $\Delta $ on the
D/R-element (Fig.~\ref{fig:Scheme}). The requirements for the wave propagation to the
backward direction are more complicated and depend both on the phase
shift $\Delta $ and on the distance $d$ between the E/A-feed and D/R
-- element.

The D/R-element fulfils a role of \emph{the director} if (i) there
is a forward constructive interference due to the phase shift $\Delta
=0$ on the D-element and (ii) a backward destructive interference due
to the phase shift $\Delta =\pi $ on the D-element and additional
phase shift on the distance $d=\lambda /4$. The D/R-element fulfils a
function of \emph{the reflector} if (i) there is a forward
destructive interference due to the phase shift $\Delta =\pi $ on the
R-element and (ii) a backward constructive interference due to the
phase shift $\Delta =0$ on the D-element and additional phase shift on
the distance $d=\lambda /4$. Unfortunately, the conditions for the
optimal phase ($\Delta =0$ or $\Delta =\pi $) are incompatible with
the condition for the strong resonance response of a D/R-element that
has the maximum at $\Delta =\pi /2$. A trade-off between the optimum
conditions for a phase and an amplitude can be found by numerical
methods.

There are too many antennas designs for us to describe them all here,
and we must concentrate on the essentials. Many of the antennas
designs share the same basic architecture, and differ mainly in the
size and number of D/R-elements. We will discuss two of the most
important D/R-elements (a wire and a sphere), and conclude with an
examination of different antennas types operating on the principles
defined by the Fano resonance.

\subsection{Thin metal wire as a radio-frequency Fano-antenna}

Let us consider the possibility of describing the far-field pattern of
radio antenna with a thin metal wire as a D/R-element by the Fano
formula. It is particularly remarkable that in the seminal paper by
Yagi \cite{yagi1928beam} one can easily identify the Fano resonance profiles in a
dependence of the current on the antennas wire length $L$.
Indeed, the emission pattern for arbitrary set of dipoles can be
represented in the form \cite{g109,PhysRevB.82.045404}
\begin{equation}\label{eq:Directivity}
c(\mathbf{n})=\sum _{i} \left((\mathbf{I}-\mathbf{n}\mathbf{n}^{T} )\mathbf{P}_{i} -\sqrt{\frac{\mu _{0}
}{\varepsilon _{0} } } \mathbf{n}\times \mathbf{M}_{i} \right) e^{-ik\mathbf{n}\mathbf{r}_{i} },
\end{equation}
where $\mathbf{P}_{i} $ and $\mathbf{M}_{i} $ are electrical and magnetic dipole moments
of an antennas elements with the position $\mathbf{r}_{i} $, $k$ is the wave
number, $\varepsilon _{0} $ and $\mu _{0} $ are the vacuum
permittivity and permeability, and $\mathbf{n}$ is a unit vector in the
direction of the radiation in the far-field. It is possible to
transform equation~(\ref{eq:Directivity}) into Fano equation~(\ref{eq:FanoFormula}) if we can separate a
frequency range in which the spectral dependence of one dipole moment
is well approximated by a resonant Lorenz function $Ae^{i\delta _{A} }
(\varepsilon -i)^{-1} $ and all others dipole moments changing slowly
and can be described by the function $Be^{i\delta _{B} } $.
Importantly, the phase difference $\delta =\delta _{A} -\delta _{B} $
which determines the Fano parameter $q$ [equation~(\ref{eq:FanoParameter})] posses a
strong directional dependence because all terms in equation~(\ref{eq:Directivity}) depend
on the direction $\mathbf{n}$.

\begin{figure*}[t]7
\includegraphics{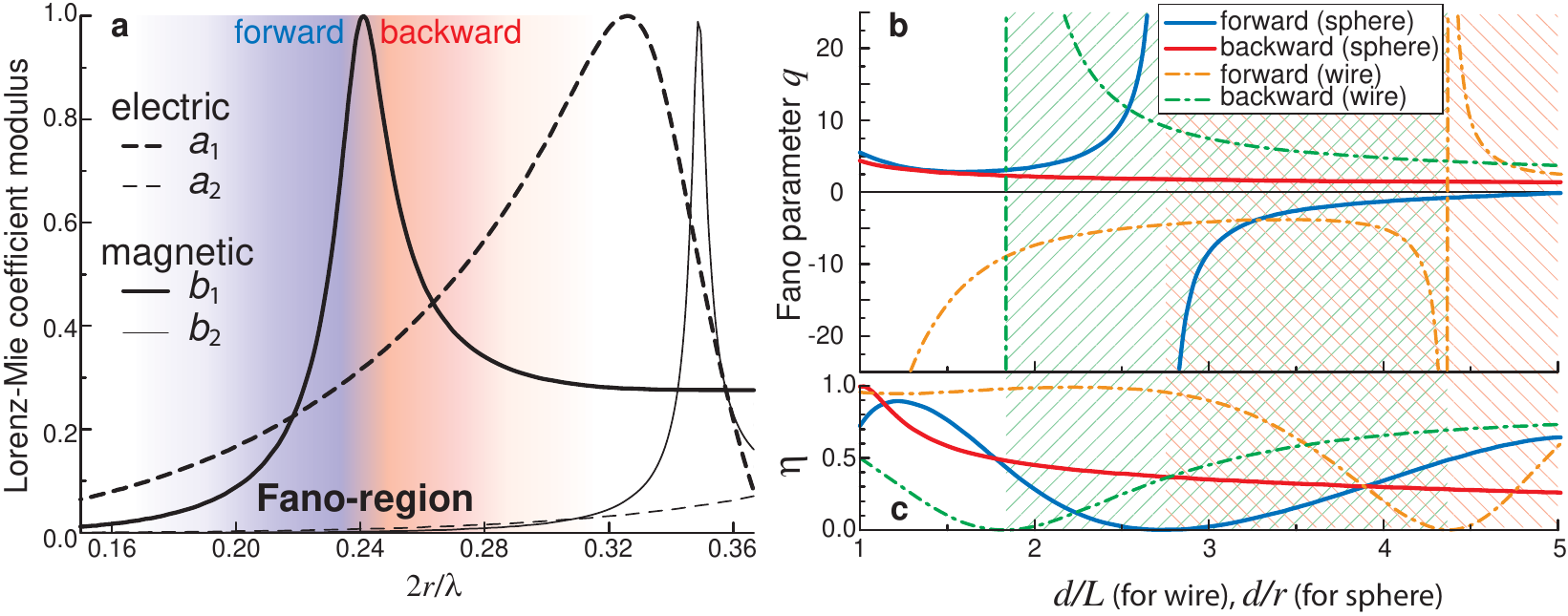}
 \caption{ (Color online) 
Fano-regimes in antennas. (a) The Lorenz-Mie
coefficient spectra of the lower multipole electric ($a_1$,
$a_2$) and magnetic ($b_1$,$b_2$) moments of a
single dielectric sphere with $\varepsilon = 16$ versus the
dimensionless frequency. In the spectral region of the Fano resonance
between isolated magnetic dipole moment $b_1$ and slowly
changing electric dipole moment $a_1$, blue indicates the region
of the forward constructive interference and red indicates the region
of the backward constructive interference in accordance with Fig.~\ref{fig:Scheme}.
(b,c) The Fano parameter $q$ and the interaction
coefficient $\eta$ as a function of the normalized distance
$d/L$ (for a wire) or $d/r$ (for a sphere). Yellow and
green shading suggest the intervals $2.75 < d/r < 5$ (for a
sphere) and $1.8 < d/L < 4.3$ (for a wire) where the Fano
parameter $q$ changes the sign for forward and backward
emission.
}
\label{fig:fanoparameter}
\end{figure*}

In the frequency range $L\approx \lambda /2$, the thin wire antenna
has an isolated electric dipole moment [Fig.~\ref{fig:fanoparameter}(a)]. The corresponding
polarizability has a resonance feature, which can be expressed as a
product of slowly varying background and the Lorenz function (see
Appendix~\ref{app:Radio}). The resonance response is specified by
a Q-factor of $\omega _{0} /\Gamma \approx 10$. If there are only two
dipoles (the E/A-feed and D/R -- element), equation~(\ref{eq:Directivity}) can be
simplified as
\begin{equation}\label{eq:DirectivityXZ}
c_{xz} (\varphi )=\cos \varphi +(a_{d} a^{e} \cos \varphi -c_{d}
a^{m} )e^{-ikd\cos \varphi },
\end{equation}
Here $\varphi$ is the angle between forward direction and the
direction $\mathbf{n}$, $a^{e} $ and $a^{m} $ are electrical and magnetic
polarizability, $a_{d} =f_{d} \left(k^{2} -d^{-2} +ikd^{-1} \right)$,
$c_{d} =f_{d} \left(k^{2} +ikd^{-1} \right)$, $f_{d} = e^{ikd}/d$.
For a thin wire antenna we have the relation $Ae^{i\delta _{A} }
(\varepsilon -i)^{-1} =a_{d} a^{e} \cos \varphi \; e^{-ikd\cos \varphi
} $ because the polarizability $a^{e} $ is proportional to the Lorenz
function, $B=\cos \varphi $ and $\delta _{B} =0$. In this case it is
possible to yield the Fano parameter $q$ from equation~(\ref{eq:FanoParameter}).

As an illustration of the approach based on the Fano resonance, we are
interesting in an antenna configuration in which one frequency
corresponds the antenna to radiate in the forward direction, and the
other frequency corresponds the antenna to radiate in the backward
direction (Fig.~\ref{fig:Scheme}). Since the Maxwell equations are scalable with
frequency and geometrical dimensions, we may reformulate the problem.
Let us fix a configuration of the Fano-antenna and find the operating
frequency for the main lobe with forward and backward orientation. The
reason is that the investigation of the spectral dependence is more
instructive and convenient for the experimental realization. Moreover,
in terms of the Fano resonance this problem has the obvious solution:
an antenna configuration should provide a different sign of the Fano
asymmetry parameter $q$ depending on the frequency of interest.

In the problem of thin wire Fano-antenna the variable parameter is the
distance $d$ between the E/A-feed and D/R -- element (Fig.~\ref{fig:Scheme}). The
dependencies of the Fano parameter $q$ and the interaction coefficient
$\eta $ on the normalized distance $d/L$ are presented in Fig.~\ref{fig:fanoparameter}(b).


\subsection{Dielectric sphere as a Fano nanoantenna}

As a D/R-element of the Fano-antenna, we can consider not only
a thin metallic wire, but also any other element having a resonant
polarizability. Here we discuss a Fano-antenna constructed from the
all-dielectric spherical particles~\cite{filonov:201113,fu2013directional}.
The advantage of dielectric materials is the ability of a direct scaling-down
of the operating frequency to the visible range as the particles size is reduced. In
contrast to the thin metallic wires, spherical particles have
both electrical and magnetic moments within the range of the magnetic resonance
for low frequencies [see Fig.~\ref{fig:fanoparameter}(a)].
Therefore, unlike the case of metallic wires, in the Fano model for
all-dielectric antennas the magnetic resonance plays the role of a narrow band.
The corresponding $Q$-factor increases almost linearly with increasing
the dielectric constant. When the dielectric constant is about 10,
the $Q$-factor is close to 10.

\begin{figure*}[t]
\includegraphics{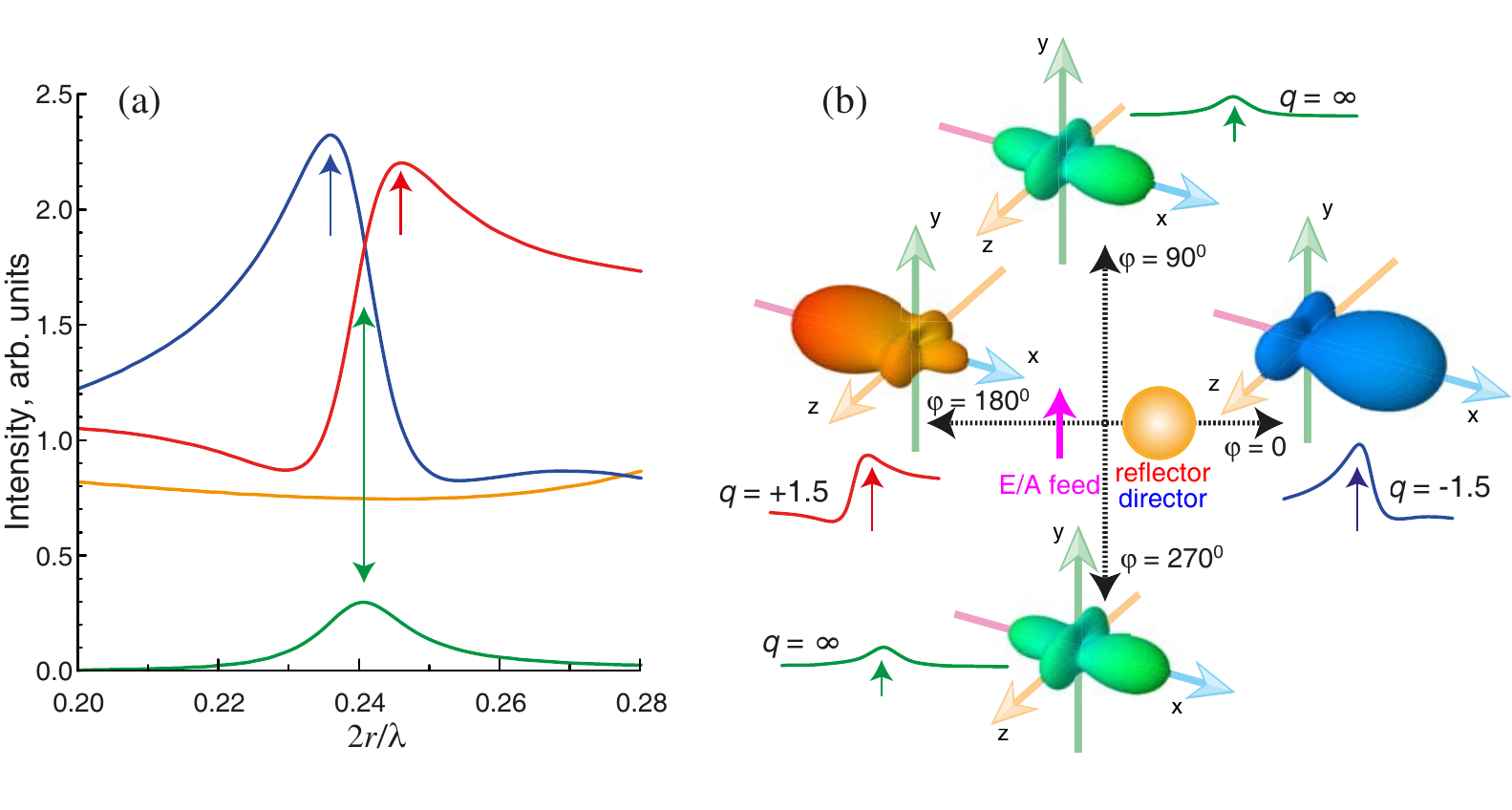}
 \caption{ (Color online) 
Spatial dependence of the Fano resonance and directivity
patterns. (a) Spectral dependencies of the far-field radiation
intensity along the $x$ axis for $\varphi =0$ (blue line) and
$\varphi =180^\circ$ (red line), along the $y$ axis, for $\varphi =90^\circ$
and $270^\circ$ (green line) and along the $z$ axis (orange line) evaluated from
Eqs.~(\ref{eq:DirectivityXZ}) and~(\ref{eq:DirectivityXY}) for a single dielectric sphere with
$\varepsilon = 16$ and $d/r=3.5$. (b) The 3D
far-field pattern for three selected frequencies marked in the panel
(a) by arrows with corresponding colors. The Fano parameter
$q$ for the main lobe is shown for each pattern.
}
\label{fig:fanoregime}
\end{figure*}

Here we will examine applicability of the Fano concept for
describing the far-field pattern of all-dielectric antennas.
We notice that the magnetic moment is pseudovector and, as a result, the
directivity pattern is formed even for zero spacing between the
E/A-feed and D/R -- element $\left(d=0\right)$. The magnetic moment is
oriented in the $z$ direction. Therefore, in addition to the
radiation pattern in the $xz$ plane, we consider the pattern
in the $xy$ plane
\begin{equation}\label{eq:DirectivityXY}
c_{xy} (\varphi )=1+(a_{r} a^{e} -c_{r} a^{m} \cos \varphi )e^{-ikd\cos \varphi }.
\end{equation}
With a large dielectric constant, the magnetic dipole resonance is
standing out against the slowly varying electric resonance [see Fig.~\ref{fig:fanoparameter}(a)].
Thus, in the $xy$ plane we have $Ae^{i\delta _{A} } (\varepsilon
-i)^{-1} =c_{r} a^{m} \cos \varphi \; e^{-ikd\cos \varphi } $ and
$Be^{i\delta _{B} } =\cos \varphi +a_{d} a^{e} \cos \varphi
e^{-ikd\cos \varphi } $; more details are given in Appendix~\ref{app:dielspheres}. For the
$xz$ plane, the similar equations have forms $Ae^{i\delta _{A} }
(\varepsilon -i)^{-1} =c_{r} a^{m} \cos \varphi e^{-ikd\cos \varphi }
$ and $Be^{i\delta _{B} } =1+a_{r} a^{e} e^{-ikd\cos \varphi } $. Now
it is easy to obtain the Fano parameter $q$ and interaction
coefficient $\eta$ from Eqs.~(\ref{eq:eta})~and~(\ref{eq:FanoParameter}). The
corresponding dependencies as functions of the normalized distance $d/r$ are
presented in Fig.~\ref{fig:fanoparameter}(b).

Figure~\ref{fig:fanoregime} summarizes our theoretical results. Intensity of the
emission spectra calculated in the dipole approximation for selected
high symmetry directions in the 3D space is presented in Fig.~\ref{fig:fanoregime}(a). The
far-field radiation intensity in the $xz$-plane for the forward
 ($\varphi =0$) and backward ($\varphi =180^\circ$)  directions is
evaluated from Eq.~(\ref{eq:DirectivityXY}). The condition when the Fano parameters for
the forward and backward directions have the same modules but different
signs ($q = \pm 1.5$) takes place at $d/r=3.5$ [Fig.~\ref{fig:fanoparameter}(b)].
Figure~\ref{fig:fanoregime} shows that at $d/r=3.5$ the maximum of the emission to the
forward direction coincide with the minimum of the emission in the
backward direction, and vice versa. The 3D far-field pattern for three
selected frequencies are marked in Fig.~\ref{fig:fanoregime}(a) by arrows with
corresponding colors. The Fano parameter $q$ for the main lobe
is shown for each pattern. The Fano-type switching of the main lobe
is illustrated in Fig.~\ref{fig:fanoregime}(b) where the 3D far-field patterns
are presented. The blue pattern (at $q = - 1.5$) demonstrates the
maximum emission in the forward direction at the frequency $2r/\lambda
= 0.236$, whereas the red pattern (at $q = + 1.5$) demonstrates the
maximum emission in the backward direction at the frequency
$2r/\lambda =0.246$.

\subsection{Experimental observation of Fano resonance in all-dielectric antennas}

\begin{figure*}[t]
\includegraphics{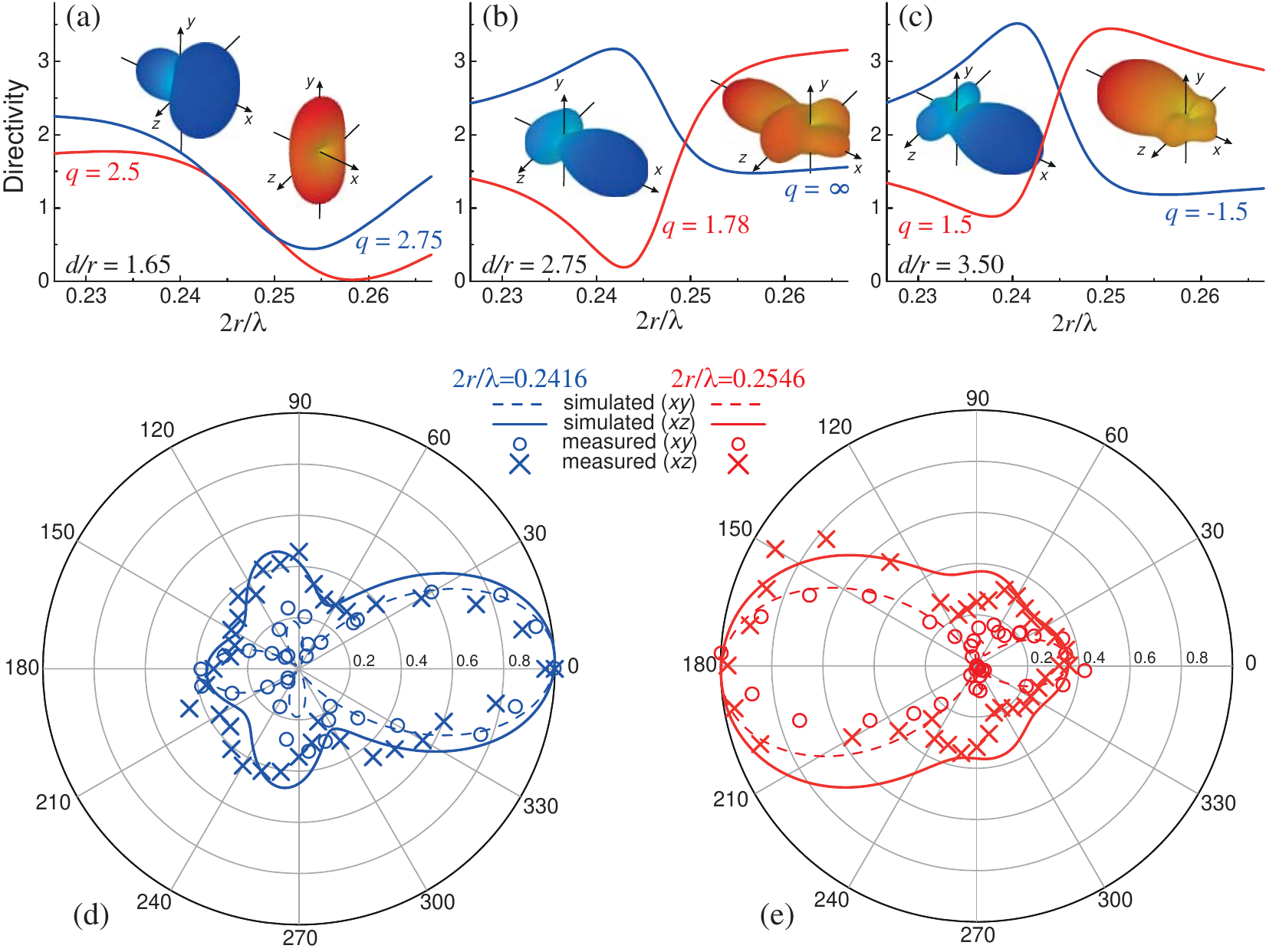}
 \caption{(Color online) 
Experimental data and numerical calculations
for the Fano effect in all-dielectric antenna. (a-c) Spectra
of forward (blue lines) and backward (red lines) directivities of the
all-dielectric antenna as a function of $d/r$ and
the corresponding Fano parameter $q$ corresponding to Fig.~\ref{fig:fanoparameter}(b). The
3D radiation patterns are presented for the frequencies 9.06 GHz
($2r / \lambda =0.2416$, blue pattern) and 9.55 GHz
($2r / \lambda =0.2546$, red pattern). The numerical results
are obtained with the CST numerical simulations.
(d,e) Radiation patterns of the
all-dielectric Fano-antenna in the $xy$ and
$xz$-planes at the frequencies 9.06 GHz (d) and 9.55~GHz
(e) for $d/r=3.5$. The symbols indicate
experimental data; solid and dashed lines show the results of
numerical CST simulations.
}
\label{fig:experiment}
\end{figure*}

To estimate the performance of the Fano-antenna, first we perform
the full wave simulations of the antenna's response by employing the CST Microwave
Studio. We consider the Fano-antenna composed of a dielectric sphere and a source.
We set the sphere permittivity and radius equal to $\varepsilon =16$
and $r = 4$~mm, respectively. As the source, we model a
short oscillator with the total length of $L = 8$~mm ($l
<\lambda/4$) and diameter of $a = 2$~mm excited by a point
source in the middle. In numerical simulations, we adjust the
distances between the sphere center and dipole to achieve an
effective switching of the main lobe [see Figs.~\ref{fig:experiment}(a-c)].
This switching occurs when the distance between sphere center and the dipole is $d
= 14$~mm ($d/r=3.5$). The corresponding radiation patterns are depicted in the
figure.

As the next step, we perform the experimental studies of the
Fano-antenna in an anechoic chamber. The dielectric sphere is made of
MgO-TiO$_{2}$ ceramic characterized by the dielectric
permittivity $\varepsilon =16$ and dielectric loss factor of
$1.12 \cdot 10^{-4} $ measured in the frequency range 8 - 12 GHz. The
sphere radius is $r = 4$~mm with the accuracy of $\pm 0.05$~mm.
To feed the vibrator with the total length of $L = 8$ mm,
we employ a coaxial cable that is connected to an Agilent E8362C
vector network analyzer. The Fano-antenna radiation patterns in the
far-field (at the distance $\cong 3$~m, $\cong 100\lambda $) is
measured by a horn antenna and rotating table. The effective switching
of the Fano-antenna main lobe measured in $xy$-and $xz$-planes has been
found at the frequencies 9.06 GHz and 9.55 GHz [see Fig.~\ref{fig:experiment}(d)]. The
measured value of the directivity is 3.5 at the frequencies 9.06 GHz
and 9.55 GHz. The measured characteristics agree very well with the
numerically simulated results.

\section{Discussion}

Our modern wireless society is based on different types of antennas
having both traditional features and advanced functionalities
including sensing, high-performance solar cells, energy communication
between nanoemitters, quantum-dot pumping etc. To meet a growing
number of new demands, an antenna should possess versatile
properties such as the ability of tuning the operating frequencies,
receiving and transmitting electromagnetic waves in different
directions. The concept of Fano-antennas provides a unique
opportunity to switch the direction of the main lobe and to
communicate at different operating frequencies.

Now we can answer positively the question that was formulated
above: using the results of our theoretical model we
demonstrate experimentally that it becomes possible to switch the
antenna's radiation direction by changing the sign of the Fano asymmetry
parameter $q$ \emph{at realistic geometric parameters of
antennas}. We present a simple analytical solution of the problem
that can be used to design a tunable Fano-antenna with required properties.
To develop a Fano-antenna, one needs to choose correctly the geometrical
parameter $d/r$ for a sphere, or the parameter $d/L$ for a wire. Specifically,
Fig.~\ref{fig:fanoparameter}(b) shows that the  parameter $q$ changes its sign
for the forward and backward emission within the intervals $2.75 < d/r < 5$ (for a
sphere) and $1.8 < d/L < 4.3$ (for a wire). According to Eq.~(\ref{eq:FanoFormula}),
for effective operation one need larger interaction coefficient $\eta$.
Therefore, it is necessary to avoid the resonance regions of $q=\pm \infty $ at $d/r=2.75$
(for a sphere) and $d/L=1.8$, 4.3 (for a wire), where
$\eta$ vanishes [see Fig.~\ref{fig:fanoparameter}(b)].

We emphasize the distinguished character of the Fano resonance
discussed here. Usually the Fano resonance is considered for isotropic
radiation or scattering and only in the spectral range without others
variable parameters~\cite{g704,francescato2012plasmonic,g908,g901,rahmani2012fano}.
In the case of Fano-antennas with the structural symmetry breaking, the  parameter
$q$ is related to the phase shifts in the forward and backward directions, and it has
a pronounced directional dependence leading to an asymmetric far-field radiation
pattern. For this reason, we introduce new variable parameter
$\varphi$ that defines the spatial radiation direction. Consequently,
the Fano parameter $q$ depends on $\varphi$ realizing \emph{the spatially dependent
Fano resonance}.

Finally, in a number of papers devoted to antennas, it is easy to
recognize the characteristic Fano-like asymmetric profiles and the
forward-backward switching effect discovered for different types of antennas,
including Yagi-Uda radio-frequency antennas~\cite{yagi1928beam},
all-dielectric antennas~\cite{filonov:201113,fu2013directional},
nonlinear hybrid metal-dielectric nanoantennas~\cite{noskov2012nonlinear}, split-ring
resonator optical antennas~\cite{jie2013antenna}, plasmonic nanoantennas~\cite{adato2011chip,rahmani2012fano,PhysRevB.82.115429,VercruysseNanoLett}.
We believe that our concept of Fano-antennas employing the parameter $\varphi$ can strongly
expand the applications of Fano resonance in nanophotonics including novel types of nanostructures
such as oligomers~\cite{francescato2012plasmonic,rahmani2012fano}), and novel functionalities such as
sensing, optical communications, light generation and routing at the nanoscale.

\section*{Acknowledgements}

The authors acknowledge useful discussions with A.E. Miroshnichenko.
This work was supported by the Ministry of Education and Science of the Russian
Federation (Projects No. 11.G34.31.0020, 14.B37.21.0307, 14.B37.21.1964 and 01201259765)
and the Government of St. Petersburg,
as well as by research grants of Dynasty Foundation (Russia), Russian Foundation
for Basic Research (project no. 11-02-00865 and 13-02-00186), the Australian Research Council
as well as a Scholarship of President of Russian Federation for young scientists
and graduate students.

\appendix

\section{Radio-frequency Fano antennas}
\label{app:Radio}

The electric polarizability of a thin-wire antenna is approximated
by the Lorenz function multiplied by a slow varying function. We solve
the inhomogeneous wave equation for the electric field by means of the
dyadic Green function~\cite{g148}. Taking into account that the tangential
component of the electric field vanishes at the antenna boundaries,
we can express the scattering field in the form
\begin{equation}
E_{y}^{scat}(\mathbf{r})=i\frac{\mu_{0}\omega}{k^{2}}(\frac{\partial^{2}}{\partial y^{2}}+k^{2})\int d\mathbf{r}'\frac{e^{ik|\mathbf{r}-\mathbf{r}'|}}{4\pi|\mathbf{r}-\mathbf{r}'|}j_{y}(\mathbf{r}').
\end{equation}
assuming that the antenna is oriented along the $y$ axis. Here $E_{y}^{scat}(\mathbf{r})$
is the $y$-component of the scattered electric field, $\omega$ is frequency,
$c$ is the speed of light, $\mu_{0}$ is magnetic constant, $k=\omega/c$
is the wavenumber, and $j_{y}$ is the current density. Using King's
approximation~\cite{g147}, the integral is evaluated as a sum of real and
imaginary functions, the latter is responsible for the radiative decay.
As a result, we find the current which allows to estimate
the electric polarizability of the antenna~\cite{g143}
\begin{equation}
a^{e}\approx\frac{4h}{\tilde{Z}_{0}k^{2}\pi}F(kh),
\end{equation}
where $F(x)$ is a function with \emph{the Lorenz function} factor
\begin{equation}
F(x)=\frac{(3x+\sin x-3iZ-8\sin\frac{x}{2})(1-\cos x)}{(3iZ-\frac{3}{2}\pi-1)}\frac{1}{x-x_{0}+i\Gamma}
\end{equation}
and the parameters of the Lorenz function are
\begin{equation}
x_{0}=\frac{\pi}{2}-\frac{4(\frac{3}{2}\pi+1)}{9Z^{2}+(\frac{3}{2}\pi+1)^{2}},
\end{equation}
\begin{equation}
\Gamma=\frac{12Z}{9Z^{2}+(\frac{3}{2}\pi+1)^{2}}.
\end{equation}
The impedance $\tilde{Z}$ reads
\begin{equation}
\tilde{Z}_{0}=\ln\left(\left(\frac{\pi h}{2a}\right)^{2}+1\right),
\end{equation}
where $a$ and $h$ is the wire radius and antenna's half-length, respectively.

\section{Dielectric spheres}
\label{app:dielspheres}

It is known that for the frequencies not exceeding the frequency of the
first electric dipole resonance, it is sufficient taking into account only the
electric and magnetic dipole moments~\cite{mulholland1994light}. The corresponding
polarizability can be found from the Lorenz-Mie coefficients $a^{e}=3i(2k^{3})^{-1}a_{1}$ and
$a^{m}=3i(2k^{3})^{-1}b_{1}$, and they have the form
\begin{equation}
a_{n}(x)=\frac{m\psi_{n}(mx)\psi_{n}'(x)-\psi_{n}(x)\psi_{n}'(mx)}{m\psi_{n}(mx)\zeta_{n}'(x)-\zeta_{n}(x)\psi_{n}'(mx)},
\end{equation}
\begin{equation}
b_{n}(x)=\frac{\psi_{n}(mx)\psi_{n}'(x)-m\psi_{n}(x)\psi_{n}'(mx)}{\psi_{n}(mx)\zeta_{n}'(x)-m\zeta_{n}(x)\psi_{n}'(mx)},
\end{equation}
presented in terms of the Riccati-Bessel functions
[real $\psi_{n}(x)$ and complex $\zeta_{n}(x)$]. Here
$m=n_{s}/n_{h}$ is the contrast of the refractive indices of the spherical
particle $n_{s}$ and surrounding medium $n_{h}$, and the size parameter
$x=kr$ is a product of the particle radius and the wavenumber.

The frequency of the first electric dipole resonance is about one and a half
times higher than the frequency of the first magnetic dipole resonance. Thus,
for sufficiently narrow resonances within the band of the magnetic dipole
resonance, the electric polarizability does not change significantly [Fig.~\ref{fig:fanoparameter}(a)].
Therefore, we can express the Lorenz-Mie coefficient $b_{1}$ as the product of \emph{the
Lorenz function and slowly varying coefficient}, namely
\begin{equation}
b_{1}(x)\approx\frac{i\gamma e^{-ix}}{m} (\sin x\psi_{1}(mx)-m\psi_{1}(x)\sin mx) \frac{1}{1-i\Omega},
\end{equation}
where
\begin{equation}
\psi_{1}(x)=\frac{\sin x}{x}-\cos x,
\end{equation}
\begin{equation}
\gamma=m-\frac{i}{\pi}(m^{2}-1),
\end{equation}
\begin{equation}
\Omega=(x-x_{0})|\gamma|^{2},
\end{equation}
\begin{equation}
x_{0}=\frac{(m^{2}-2)(m^{2}-1)+\pi^{2}m^{2}}{m\pi|\gamma|^{2}},
\end{equation}
\begin{equation}
\Gamma=\frac{2}{|\gamma|^{2}}.
\end{equation}


\end{document}